\documentclass[onecolumn]{emulateapj}
\usepackage[hyperfootnotes=false,colorlinks=true,linkcolor=blue,citecolor=blue]{hyperref}
\slugcomment{To appear in ApJ 644, June 20, 2006}
\shortauthors{Thommes \& Murray} \shorttitle{Surviving Type I Migration}

\begin{document}

\title{Giant planet accretion and migration: Surviving the Type I regime}
\author{Edward W. Thommes and Norman Murray}
\affil{Canadian Institute for Theoretical Astrophysics, University of Toronto, 60 St. George Street, Toronto, Ontario M5S 3H8}
\email{thommes@cita.utoronto.ca}

\begin{abstract}
In the standard model of gas giant planet formation, a large solid core ($\sim 10$ times the Earth's mass) forms first, then accretes its massive envelope (100 or more Earth masses) of gas.  However, inward planet migration due to gravitational interaction with the proto-stellar gas disk poses a difficulty in this model.  Core-sized bodies undergo rapid ``Type I" migration; for typical parameters their migration timescale is much shorter than their accretion timescale.  How, then, do growing cores avoid spiraling into the central star before they ever get the chance to become gas giants?
Here, we present a simple model of core formation in a gas disk which is viscously evolving.  As the disk dissipates, accretion and migration timescales eventually become comparable.  If this happens while there is still enough gas left in the disk to supply a jovian atmosphere, then a window of opportunity for gas giant formation opens.  We examine under what circumstances this happens, and thus, what predictions our model makes about the link between proto-stellar disk properties and the likelihood of forming giant planets.  
\end{abstract}

\keywords{protoplanetary disks, origin: solar system --- planets:
formation, planets:extrasolar}

\section{Introduction}
\label{sec: intro}
The observed correlation of planet occurrence and host star metallicity \citep{1997MNRAS.285..403G,2005ApJS..159..141V},
together with the lack of evidence \citep{2002AJ....124..400Q,2003ApJ...596L.101D,2005ApJ...622.1102F}
 for accretional pollution of stellar convective zones 
 \citep{1997ApJ...491L..51L,2001ApJ...555..801M}
 implies that solids in the disk aid giant planet forrmation.  This in turn provides support for the core-accretion model of giant planet formation, wherein a large protoplanet, $\sim 10$ Earth masses (M$_\oplus$), forms first, then accretes a massive gas envelope
\citep{1978PThPh..60..699M,1996Icar..124...62P}.
 The growth of such a large body in itself provides a challenge to our understanding of planet accretion.  It may require some or all of the following: large disk masses 
 \citep{1987Icar...69..249L,2003Icar..161..431T},
 local enhancements in solids surface density 
 \citep{1988Icar...75..146S,2004ApJ...614..490C}
  and a substantial fraction of planetesimals at $\ll$ km size in order to increase the effectiveness of aerodynamic gas drag, thus increasing the capture cross-section of the core atmosphere 
\citep{2003Icar..166...46I}
and making planetesimal random velocities very small
\citep{2004AJ....128.1348R}.
However, a further problem plagues the core accretion model.  Gravitational interaction with the parent gas disk causes the inward migration of embedded planetary bodies 
\citep{1997Icar..126..261W,2000MNRAS.315..823P,2002ApJ...565.1257T};
 this is referred to as Type I migration.  The migration rate increases with planet mass until the planet becomes large enough, of order $10^2$ M$_\oplus$, to open a gap in the disk 
 \citep{1980ApJ...241..425G,1986ApJ...309..846L},
  thus switching to what is called Type II migration.  The migration rate is therefore peaked in the mass range of cores, of order $10^1$ M$_\oplus$, and for most disk models far outstrips the accretion rate.  In order to prevent bodies from falling into the star long before they reach core mass, studies of concurrent growth and migration of giant planets have thus far resorted to assuming a large reduction in the Type I rate 
  \citep{2005A&A...434..343A},
or setting it to zero 
\citep{2002A&A...394..241T,2004ApJ...616..567I}.
  Though arguments can be made for doing so (see \S \ref{sec: discussion}), it is nevertheless troubling if this is really the only way to salvage core accretion.  Thus, our purpose here is to address the question of whether unmitigated Type I migration really does doom giant planet cores.  To do this, we construct a simple semi-analytic model of concurrent core accretion and Type I migration in a viscously evolving protostellar disk.  We find that although cores cannot form in a young disk, the situation changes as the disk dissipates.  We conclude that, depending on disk properties, there exists a limited timespan, after the disk becomes sufficiently tenuous but before it can no longer supply a jovian atmosphere, within which giant planet cores can successfully grow.

\section{Growth and migration rates}
\label{sec: rates}

Since we want to concurrently calculate the migration and accretional growth of protoplanets, we require analytic expressions for both.  For the former, we use the result of 
\cite{2002ApJ...565.1257T}:
\begin{equation}
\dot{r}_{\rm migr} = \left (2.7 + 1.1 \beta \right ) \frac{M}{M_*}\frac{\Sigma_g r^2}{M_*} \left ( \frac{r \Omega}{c_s}\right )^2 \Omega r
\label{eq: Type I rate}
\end{equation}
for a mass $M$ body orbiting a mass $M_*$ star at radius $r$, embedded in a gas disk with surface density $\Sigma_g$ and sound speed $c_s$.  $\Omega$ is the Keplerian angular velocity and $\beta=-d \log \Sigma_g/d \log r$.
For the latter, we assume ``oligarchic'' growth 
\citep{1998Icar..131..171K,2000Icar..143...15K}
and make use of the corresponding growth rate estimate of \cite{2003Icar..161..431T}.
For simplicity, we neglect the radial motion of planetesimals due to aerodynamic gas drag, the enhanced capture radius of growing protoplanets due to formation of an atmosphere, and fragmentation of planetesimals.  
The growth rate is then
\begin{equation}
\left . \frac{D M}{D t} \right |_{\rm accr} \approx A \Sigma_{m} M^{2/3},
\label{eq: oligarchic growthrate}
\end{equation}
where
\begin{equation}
A =5.9 \frac{(C_{D} \,b\, \rho_{g})^{2/5} G^{1/2} M_{*}^{1/6}}{\rho_{m}^{4/15} \rho_{M}^{1/3} r^{1/10} m^{2/15}}.
\label{eq: oligarchy A factor}
\end{equation}
Here $\Sigma_{m}$ is the surface density of planetesimals, $M$ is the protoplanet mass, $m$ is the planetesimal mass, $\rho_M$ and $\rho_m$ are the densities of a protoplanet and a planetesimal, $C_{D}$ is a dimensionless drag coefficient $\sim 1$ for km-sized or larger planetesimals, and $b$ is the spacing between adjacent protoplanets in units of their Hill radii.  An equilibrium between mutual gravitational scattering on the one hand and dynamical friction from the planetesimals on the other keeps $b \sim 10$ \citep{1998Icar..131..171K}.  

The derivation of Equation \ref{eq: oligarchic growthrate} assumes for simplicity that planetesimal random velocities are always at their equilibrium value.  
\cite{1999Icar..139..350T} show that rapid migration can significantly boost the planetesimal accretion rate of a protoplanet by allowing it to accrete planetesimals with low random velocities existing further inward in the disk.  However, this assumes that the inner planetesimal disk is pristine.  In reality, the shorter accretion timescale at smaller radius means that inward migration delivers a protoplanet into a region that has already been gravitationally ``pre-stirred" by earlier protoplanets of comparable size.  Thus we consider our simplification reasonable, and leave time-dependent calculation of planetesimal random velocities as a future refinement to the model.  More importantly, in addition to pre-stirring, these earlier protoplanets have pre-depleted the inner disk.  To capture this effect, we track the global evolution of the planetesimal disk (Equation \ref{Sigma PDE} below).

Another assumption which goes into Equation \ref{eq: oligarchic growthrate} is that protoplanets do not open gaps in the planetesimal disk.  Since the planetesimal disk has a much lower effective viscosity than the gas disk, a single isolated protoplanet will open a gap early on in its growth, unless its migration time is less than its gap-opening time 
\citep{1995ApJ...440L..25W,1999Icar..139..350T}.
However, when multiple protoplanets exist in close proximity, planetesimal disk gaps cannot form
\citep{1997Icar..125..302T,2003Icar..161..431T}.

The net rate of change of the protoplanet mass at a given radius $r$ in the disk (the Eulerian derivative of $M$) is then the protoplanet growth rate (the Lagrangian derivative of $M$) plus the mass rate of change due to migration (the advection term):
\begin{equation}
\frac{\partial M(r,t)}{\partial t} =  \left . \frac{D M}{D t} \right |_{\rm accr} - \dot{r}_{\rm migr} \frac{\partial M}{\partial r}.
\label{M PDE}
\end{equation}
\cite{2006Icar..180..496C} points out that the maintenance of a constant spacing in Hill radius between adjacent protoplanets implies that  protoplanets must merge, at a rate constituting $50 \%$ of the mass accretion rate due to planetesimals.  This result is obtained for protoplanets that are radially fixed; with migration, protoplanet orbits gradually diverge (as can be seen by skipping ahead to Figure \ref{fig: timeline_projections}), so that less merging between neighbors is required in order to keep a constant Hill spacing.  Nevertheless, in neglecting protoplanet-protoplanet mergers, we are underestimating the protoplanet growth rate.  We leave a detailed examination of this issue to future work.  

As the protoplanets grow, they deplete the surface density of planetesimals in the nearby part of the disk.  If a given protoplanet accretes from an annulus of width $\Delta r$, then the rate of change of total planetesimal mass in that annulus is given by
\begin{equation}
\frac{d}{dt}  \left ( 2 \pi r \Delta r \Sigma_{m} \right ) = -\left . \frac{D M}{D t} \right |_{\rm accr} 
\end{equation}
Using $\Delta r = b r_{H}$, we obtain 
from this the surface density rate of change in terms of the protoplanet accretion rate (since we neglect planetesimal migration, this equation contains no advection term):
\begin{equation}
\frac{\partial \Sigma_{m}(r,t)}{\partial t}=-\frac{1}{3^{2/3} b \pi r^{2}} \left ( \frac{M_{*}}{M} \right )^{1/3}\left . \frac{D M}{D t} \right |_{\rm accr} 
\label{Sigma PDE}
\end{equation}

We can then solve the system of partial differential equations (\ref{M PDE}) and (\ref{Sigma PDE}) to obtain $M(r,t)$ and $\Sigma_{m}(r,t)$.  

\section{The disk model}
\label{sec: disk model}
In order to model the viscous evolution of the gas disk in a simple way, we make use of the similarity solution of
\cite{1974MNRAS.168..603L}
We assume an $\alpha$-parameterization of viscosity, $\nu \equiv \alpha c_s H$, where $c_s$ is the sound speed and $H \approx c_s/\Omega$ is the gas disk scale height, and assume $\alpha$ to be constant.  We adopt a disk temperature profile $T \propto r^{-1/2}$ (i.e. $c_s \propto r^{-/4}$).  The similarity solution is then of the form
\citep{1998ApJ...495..385H}
\begin{equation}
\Sigma_{\rm g}(r,t)=\frac{M_{d}(0)}{2 \pi R_{0}^{2}} \frac{1}{(r/R_{0})\tau_{d}^{3/2}}e^{-(r/R_{0})/\tau_{d}}.
\label{eq: similarity disk}
\end{equation}
Here $M_{d}(0)$ is the total disk mass at $t=0$, $R_{0}$ is the characteristic disk radius at $t=0$ (beyond which surface density falls off exponentially), and $\tau_{d} \equiv t/(R_{0}^{2}/3 \nu_0)+1$ is the nondimensional time.  For $r << R_0 \tau_d$, $\beta$ (Equation \ref{eq: Type I rate}) $\approx 1$.  

We then scale the surface density of the planetesimal disk to that of the gas disk at $t=0$, adding in a factor of $0.24$ decrease inside the water evaporation radius, for which we adopt 2.7 AU as in the model of 
\cite{1981PThPh..70...35H}:
\begin{equation}
\Sigma_{\rm m}(r,0) = 0.019 \times10^{\rm [Fe/H]} f \Sigma_{\rm g}(r,0)
\end{equation}
where [Fe/H] is the metallicity index (0=Solar abundance).  We choose [Fe/H]=0.25.  The factor $f$ introduces the surface density jump at the snow line, smoothed over a radial scale of 1 AU:
\begin{equation}
f=0.24+(1-0.24) \left [ \frac{1}{2}\tanh \left (\frac{r-2.7\,{\rm AU}}{\rm 1\,AU} \right ) +\frac{1}{2} \right ]
\end{equation}
For the sound speed, we use $c_{s}=1.4 \times 10^{5}(r/{\rm AU})^{-1/4}\,{\rm cm\,s^{-1}}$ which yields a disk scale height $H \approx 0.047(r/{\rm 1\,AU})^{5/4}$\,AU, again as in the model of 
\cite{1981PThPh..70...35H}.
  In addition, we adopt $\rho_{m}=\rho_{M}=1.5{\rm \,g\,cm^{-3}}$, $b=10$, and $m=10^{-12}$ M$_{\oplus}$ (which corresponds to $r_{m} \approx$ 1 km).

\section{Comparison of timescales}
\label{sec: timescales}
Before we proceed with solving Equations \ref{M PDE} and \ref{Sigma PDE}, we can can gain insight into the problem of concurrent protoplanet accretion and migration from a simple analysis of the timescales involved.  The accretion and migration timescales of a protoplanet are
\begin{equation}
t_{\rm accr} \equiv \frac{M}{D M/\left .D t \right |_{\rm accr} }
\label{eq: accretion timescale}
\end{equation}
and
\begin{equation}
t_{\rm migr} \equiv \frac{r}{ \dot{r}_{\rm migr} }.
\label{eq: migration timescale}
\end{equation} 
Adopting a disk mass $M_d=0.15$ M$_\odot$ and radius $R_0=50$ AU, the model gas disk of \S \ref{sec: disk model} above has $\Sigma_g({\rm 5\,AU})\approx 770\,{\rm g\,cm^{-2}}$ at $t=0$, and a midplane gas volume density $\rho_g\sim \Sigma_g/2 H$, where $H({\rm 5\,AU})\approx 0.07$ AU.  We can write the accretion and migration timescales as 
\begin{eqnarray}
t_{\rm accr} & \sim & 2.4 \times 10^5 \left ( \frac{\Sigma_g}{\rm 770\,g\,cm^{-2}}\right )^{-2/5} \left ( \frac{H}{0.07 r} \right )^{2/5} \left ( \frac{r}{\rm 5\,AU} \right )^{1/2} \left ( \frac{\Sigma_m}{0.019 \times 770 \times 10^{0.25} {\rm g\,cm^{-2}}} \right )^{-1} \nonumber\\
\, & \, & \times \left ( \frac{M}{\rm 10\,M_\oplus}\right )^{1/3}\,{\rm yrs},\nonumber\\
t_{\rm migr} & \sim & 3.6 \times 10^4 \left ( \frac{\Sigma_g}{\rm 770\,g\,cm^{-2}} \right )^{-1} \left (\frac{H/r}{0.07} \right )^2 \left (\frac{r}{\rm 5\,AU} \right )^{3/2} \left (\frac{M}{\rm 10\,M_\oplus} \right )^{-1}\,{\rm yrs}
\label{eq: timescales with numbers}
\end{eqnarray}
Thus the accretion timescale of a 10 M$_\oplus$ body at 5 AU starts out $\approx 6.7 \times$ as long as its migration timescale, meaning that a protoplanet will fall into the star well before it reaches core size.  However, the accretion time has a weaker dependence on gas density than does the migration time; $t_{\rm accr} \propto \Sigma_g^{-2/5}$ while $t_{\rm migr} \propto \Sigma_g^{-1}$.  Therefore if the gas density is reduced far enough, accretion can win out over migration.  In particular, for the above example, reducing $\Sigma_g$ by a factor $6.7^{-5/3}=0.04$ will make the two timescales equal.  

\begin{figure}
\epsscale{0.8}
\plotone{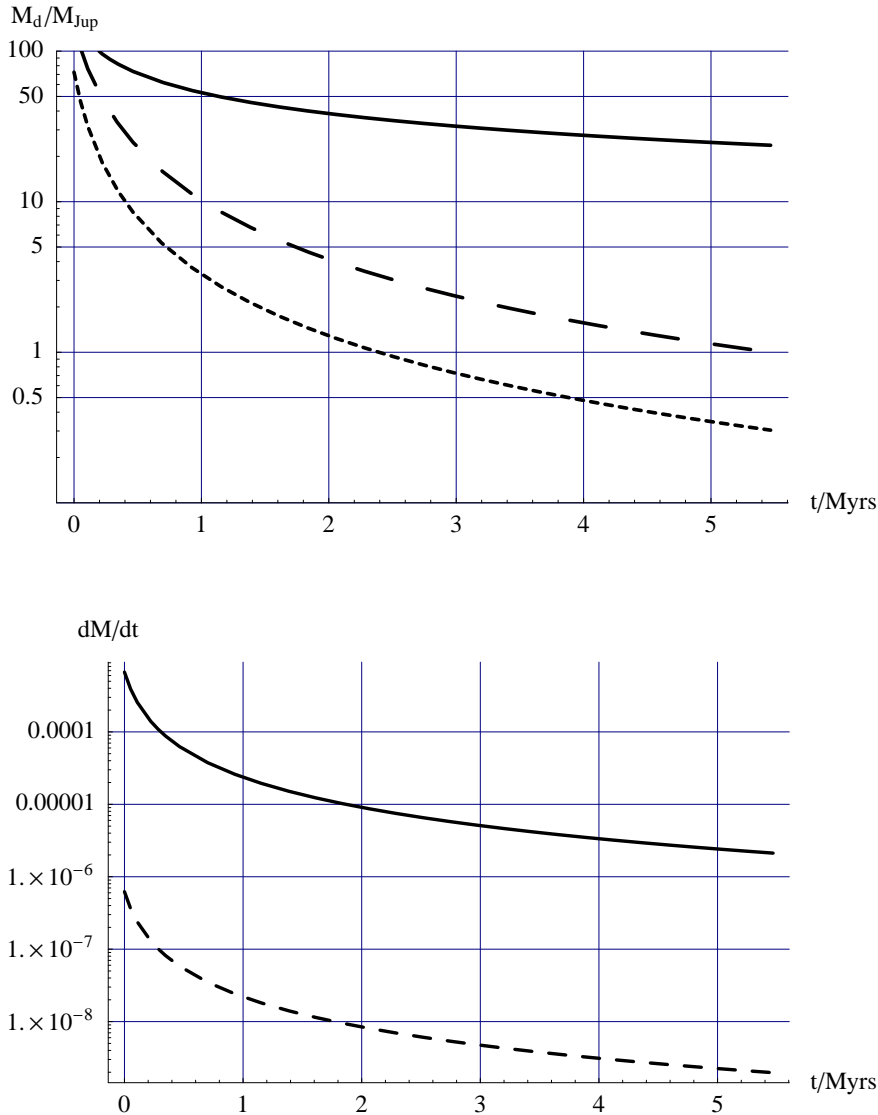}
\caption{Time evolution of an $\alpha$ disk having the sound speed profile of \S \ref{sec: disk model} and $\alpha=10^{-2}$, initial disk mass $M_d(0)=0.15$ M$_\odot$, and radius $R_0=50$ AU.  TOP: The total gas mass ({\it solid}), the gas mass inside 100 AU ({\it long-dashed}), and the gas mass inside 30 AU ({\it short-dashed}).  BOTTOM: The mass flux in the inner disk in units of M$_{\rm Jup}$/yr ({\it solid}) and M$_\odot$/yr ({\it dashed}).}
\label{fig: baseline_gas}
\epsscale{1}
\end{figure}

We can also construct a more quantitative estimate.  In the absence of migration, we would estimate the protoplanet mass at a time $t_1$ by setting $t_1=t_{\rm accr}$, then using this together with Equations \ref{eq: oligarchic growthrate} and \ref{eq: accretion timescale} to solve for $M_{\rm accr}(r,t_1)$.  The upper bound on accretion is the oligarchic isolation mass, i.e. the mass at which a protoplanet has consumed all planetesimals within its feeding zone, $\Delta r = b r_H$:
\begin{equation}
M_{\rm iso}(r) = 2 r^3\sqrt{\frac{2 b^3 \pi^3 \Sigma_m^3}{3 M_*}}.
\end{equation}
With the protoplanets migrating, we can estimate an upper bound on accretion by computing the mass $M_{\rm cross}(r)$ at which $t_{\rm accr}=t_{\rm migr}$.  It then remains to pick a limiting time, after which the disk no longer contains enough gas to provide a jovian atmosphere.  Using the disk model of \S \ref{sec: disk model} with $\alpha=10^{-2}$ and, as above,  $M_d=$ 0.15 M$_\odot$ and $R_0=$ 50 AU, we examine the time evolution (Figure \ref{fig: baseline_gas}).  We compute the times at which the disk has dissipated to the point that there is a Jupiter mass of gas left inside 30 AU and 100 AU: $t_{\rm 30\,AU}=2.4$ Myrs, $t_{\rm 100\,AU}=5.5$ Myrs.  For comparison, the {\it total} disk mass at these times, given by
$M_d(0) \tau_d^{-1/2}$, is 35 M$_{\rm Jup}$ and 24 M$_{\rm Jup}$, respectively.  Also, the gas accretion rate onto the star---and thus the approximate mass flux anywhere in the inner disk---is given by $\dot{M_d}=(3/2) M_d(0) (\nu_0/R_0^2) \tau_d^{-3/2}$, and is $7.0 \times 10^{-6}$ M$_{\rm Jup}$/yr and $2.1 \times 10^{-6}$ M$_{\rm Jup}$/yr, respectively.  This suggests $t_{\rm 30\,AU}$ as well as $t_{\rm 100\,AU}$ constitute reasonable upper limits on how fast a core must form in order to still be able to accrete a gas giant atmosphere from the disk.

\begin{figure}
\epsscale{1}
\plotone{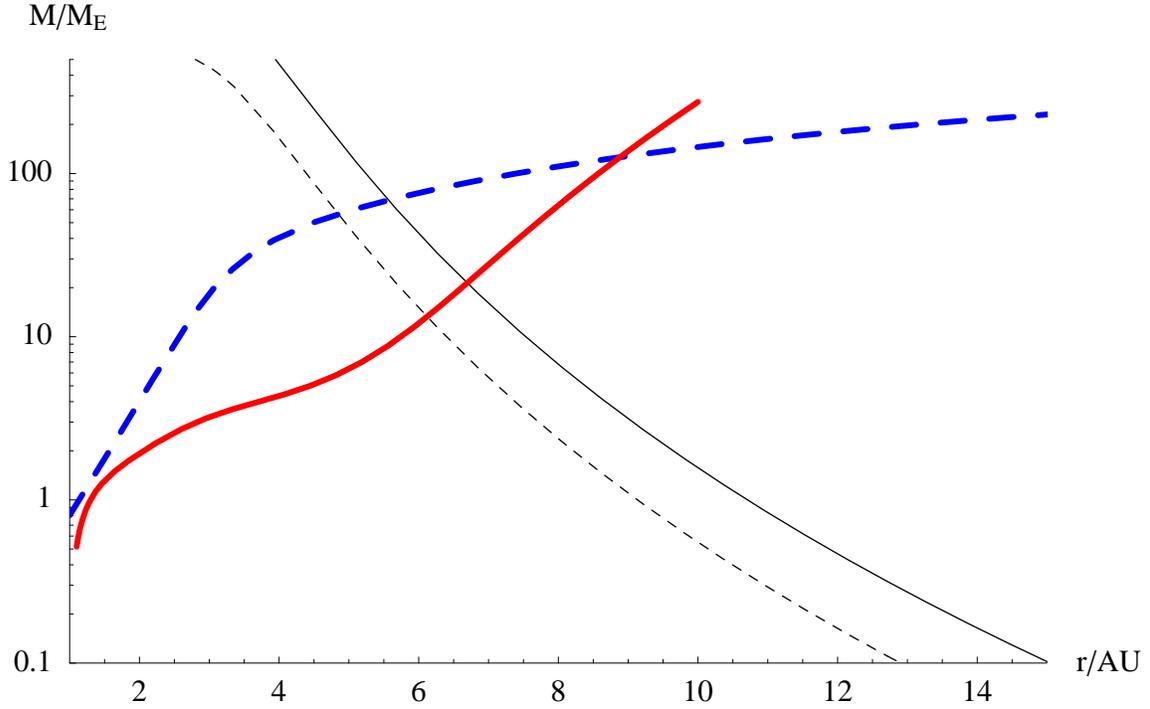}
\caption{Isolation mass $M_{\rm iso}(r)$ {\it (thick dashed curve)} and ``crossover" mass $M_{\rm cross}(r)$ {\it (thick solid curve)} at which $t_{\rm accr}=t_{\rm migr}$, together with $M_{\rm accr}(r,t_{\rm 30\,AU})$ {\it (thin dashed curve)} and $M_{\rm accr}(r,t_{\rm 100\,AU})$ {\it(thin solid curve)} for the gas disk in Figure \ref{fig: baseline_gas}, with all other parameters as in \S \ref{sec: disk model}.  At $t=t_{\rm 30\,AU}$, our estimate of the protoplanet mass as a function of {\it original} stellocentric radius is $M(r)=\min(M_{\rm accr}(r,t_{\rm 30\,AU}),M_{\rm iso}(r),M_{\rm cross}(r))$; at $t=t_{\rm 100\,AU}$, the estimate is $M(r)=\min(M_{\rm accr}(r,t_{\rm 100\,AU}),M_{\rm iso}(r),M_{\rm cross}(r))$}
\label{fig: 4line_estimate_plot}
\end{figure}

Figure \ref{fig: 4line_estimate_plot} shows a plot of $M_{\rm iso}(r)$ and $M_{\rm cross}(r)$, together with $M_{\rm accr}(r,t_{\rm 30\,AU})$ and $M_{\rm accr}(r,t_{\rm 100\,AU})$.     In the absence of migration, our estimate of the protoplanet mass as a function of radius at time $t$ would be 
$$
M(r,t)=\min(M_{\rm accr}(r,t),M_{\rm iso}(r)).
$$
With migration, we can estimate the protoplanet mass as a function of (initial) orbital radius as
$$
M(r,t)=\min(M_{\rm accr}(r,t),M_{\rm iso}(r),M_{\rm cross}(r)).
$$  
For comparison with the subsequent full computation, in which we use 1 AU as the inner boundary, we take for $t_{\rm migr}$ the time to migrate to 1 AU, rather than all the way to the star.
To estimate the {\it largest} protoplanet mass at time $t$, we find the masses at which $M_{\rm accr}(r,t)$ intersects $M_{\rm iso}(r)$ and $M_{\rm cross}(r)$, and pick the smaller of the two:
\begin{equation}
M_{\rm max}(t)=\min \left ( M_{\rm accr}(r,t) \cap M_{\rm iso}(r), M_{\rm accr}(r,t) \cap M_{\rm cross}(r) \right )
\label{eq: maximum mass estimate}
\end{equation}

In this way, we estimate the largest protoplanet mass at $t_{\rm 30\,AU}$ and  $t_{\rm 100\,AU}$ to be $\approx$ 10 M$_\oplus$ and $\approx$ 20 M$_\oplus$,  respectively.  Thus, we predict that this set of parameters will indeed produce core-sized bodies, and do so while there is still enough gas in the disk to plausibly provide a gas giant envelope.  

\section{Computation of the model}
\label{sec: computation}

{\subsection{The simplest case: No accretion, no gas disk evolution}
\label{sec: simplest}
We now proceed with a full solution of Equations (\ref{M PDE}) and (\ref{Sigma PDE}).  The equations are solved numerically using Mathematica 5 and  MATLAB 7.  We take 1 AU as the inner boundary of our computation domain.
For comparison, we begin by neglecting Type I migration altogether, and we fix the gas disk at its $t=0$ value.  
The result of the calculation is shown in Figure \ref{fig: no_migr_no_disk_evol}, for a disk mass of $M_d=0.15$ M$_\odot$.  
\begin{figure}
\plotone{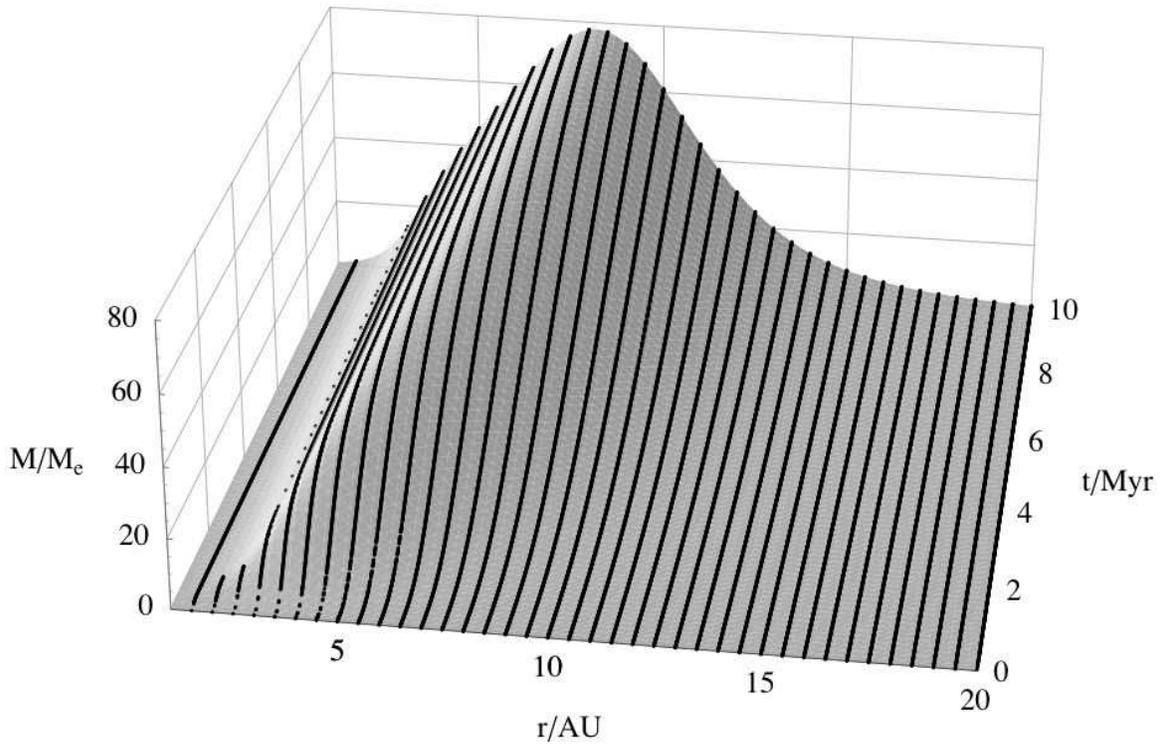}
\caption{Protoplanet mass as a function of stellocentric radius and time for the case of \S \ref{sec: simplest}:  A time-invariant gas disk of mass $M_d=0.15$ M$_\odot$, radius $R_0=50$ AU, metallicity [Fe/H]=0.25, and no Type I migration.  The overplotted streamlines show examples of protoplanet evolution paths; because there is no migration, all are parallel to the $t$-axis.}
\label{fig: no_migr_no_disk_evol}
\end{figure}
A largest protoplanet mass of 10 M$_\oplus$ is reached in $\approx 2$ Myrs.  After 10 Myrs, the largest protoplanet mass is nearly 80 M$_\oplus$, with this maximum occurring between 5 and 10 AU.  Thus, in the absence of migration, this protostellar disk readily and rapidly produces bodies of more than sufficient mass to furnish jovian cores.   

\subsection{Adding migration}
\label{sec: adding migration}
Next, we include Type I migration (Equation \ref{eq: Type I rate}) and repeat the calculation.  Results are shown in the top panel of Figure \ref{fig: migr_no_disk_evol}.
\begin{figure}
\epsscale{0.75}
\plotone{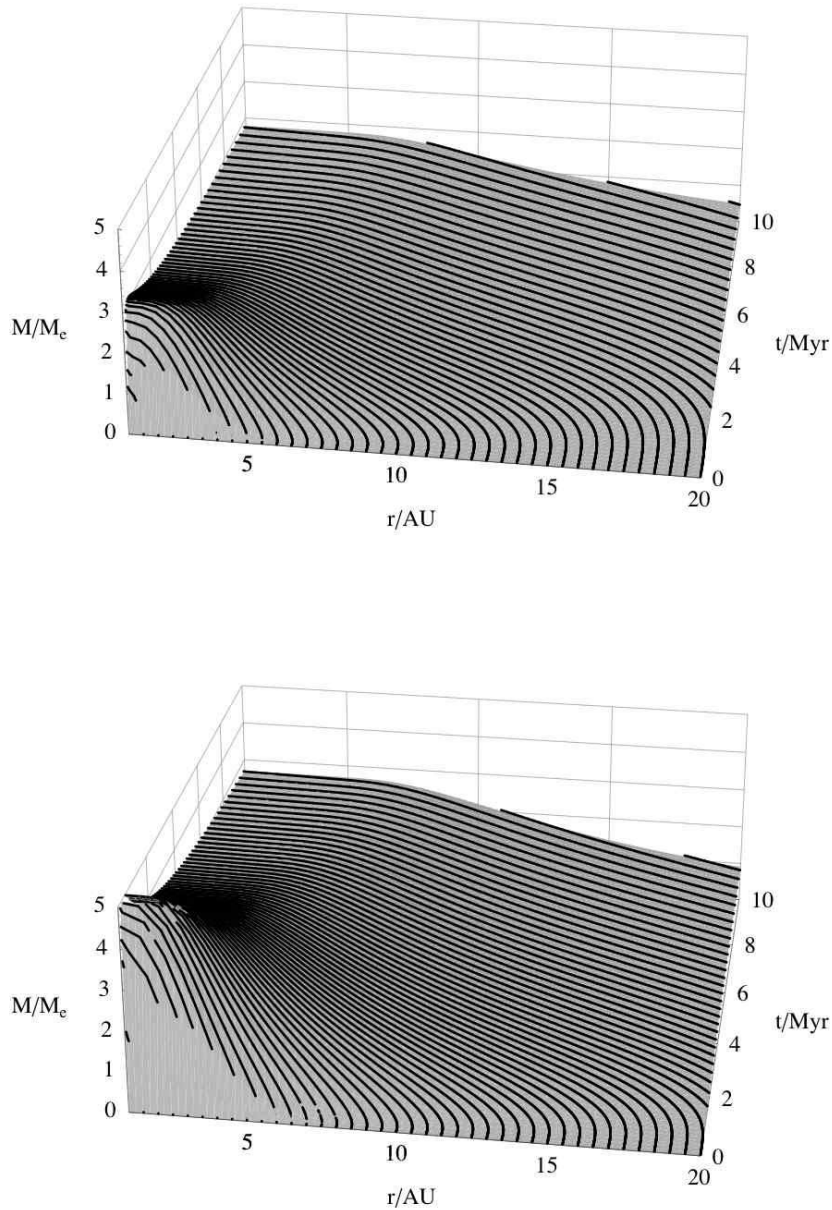}
\caption{Protoplanet mass as a function of stellocentric distance and time with Type I migration,  in a time-invariant gas disk  of mass $M_d=0.15$ M$_\odot$, radius $R_0=50$ AU (\S \ref{sec: adding migration}).  The overplotted streamlines show examples of protoplanet evolution paths.  The initial spacing between streamlines is 0.5 AU, and does not represent the actual radial distance between adjacent protoplanets.  {\it TOP:} Disk with [Fe/H]=0.25; {\it BOTTOM:} [Fe/H]=0.5}
\label{fig: migr_no_disk_evol}
\epsscale{1}
\end{figure}
The outcome is now dramatically different.  The largest protoplanet mass reached is just over 3 M$_\oplus$; this occurs at the inner boundary within the first million years.  After 10 Myrs, the largest protoplanet mass is just under  2 M$_\oplus$.  The overplotted protoplanet evolution paths show that Type I migration removes a prodigious amount of material:  By $\sim 5$ Myrs, all the protoplanets originating inside 20 AU have fallen through the inner boundary at 1 AU.  We repeat the calculation once more, now increasing the metallicity to [Fe/H]=0.5.  This makes the planetesimal surface density 46 g/cm$^2$ at 5 AU, about $17 \times$ the value at that radius in the minimum-mass Solar nebula model.
\citep{1981PThPh..70...35H}.
  The outcome is plotted in the bottom panel of Figure \ref{fig: migr_no_disk_evol}.  The mass after 10 Myrs is still less than $3$ M$_\oplus$, though it briefly exceeds 5 M$_\oplus$ at the inner edge within the first million years.  This illustrates how in a time-invariant gas disk, even a very high-metallicity one,
Type I migration completely overpowers accretion.  

\subsection{Adding viscous evolution of the gas disk}
\label{sec: baseline}

Dropping the metallicity back down to [Fe/H]=0.25, we now perform the calculation with the gas disk evolving in time according to Equation \ref{eq: similarity disk}.  As in \S \ref{sec: timescales}, we chose $\alpha=10^{-2}$, $M_d=0.15$ M$_\odot$, and $R_0=50$ AU.  The result is shown in Figure \ref{fig: baseline_full_calculation}.
\begin{figure}
\epsscale{1}
\plotone{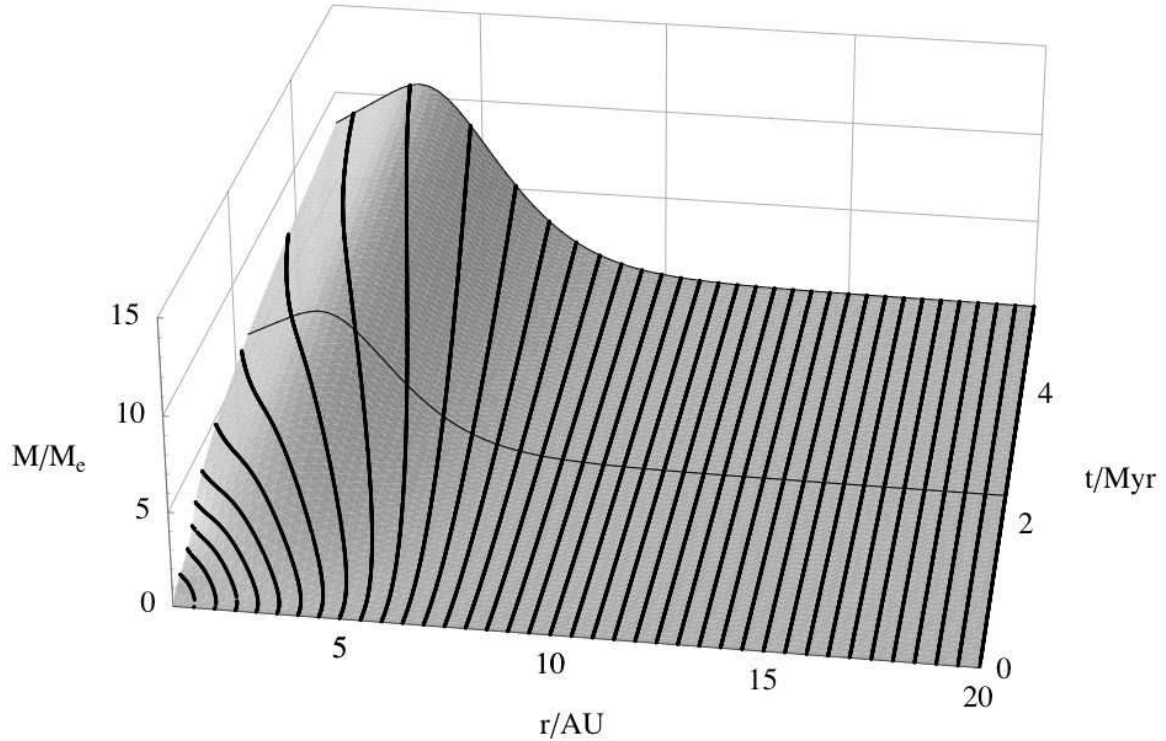}
\caption{Protoplanet mass as a function of stellocentric distance and time with Type I migration, in an evolving $\alpha$ disk having $\alpha=10^{-2}$, $M_d=0.15$ M$_\odot$, $R_0=50$ AU and [Fe/H]=0.25 (\S \ref{sec: baseline}).  The gas disk evolution is thus the same as that shown in Figure \ref{fig: baseline_gas}.  The overplotted streamlines show examples of protoplanet evolution paths.  The calculation is carried forward to $t_{\rm 100 AU}$=5.5 Myrs (see \S \ref{sec: timescales}), and a line is drawn showing $t_{\rm 30 AU}$=2.4 Myrs.  }
\label{fig: baseline_full_calculation}
\epsscale{1}
\end{figure}

The computation is carried forward to $t=t_{\rm 100\,AU}=5.5$ Myrs.  By this time, the largest protoplanet mass reached is just over 10 M$_\oplus$, around 3 AU.  Before this, at $t_{\rm 30\,AU}=2.4$ Myrs, the maximum protoplanet mass is $\approx 7$ M$_\oplus$.  The masses produced are thus comparable to (though somewhat lower than) the estimates from \S \ref{sec: timescales}; the model succeeds in producing bodies large enough to plausibly serve as giant planet cores while there is still enough gas in the disk to furnish a jovian-mass atmosphere.  

Looking at the overplotted protoplanet evolution paths, we see that only the protoplanets originating at $\la 5$ AU are lost.  It is the short growth timescale at small $r$ which dooms them; they become massive too early, whereas the ones slightly further out remain small until the gas disk surface density has dropped to a safer level.  

Figure \ref{fig: timeline_projections} shows another view of the protoplanets' time evolution, projecting the evolution streamlines into the $t-r$ and $t-M$ planes.  This gives an even clearer view of the range of outcomes:  The promising jovian core candidates, which attain a mass close to 10 M$_\oplus$ by $t_{\rm 100\,AU}$, originate from narrow radial range, about 5 to 6.5 AU.  The ones starting further in are lost before they grow massive enough, while the ones starting further out simply don't do much accreting at all in the time available.  Also visible is the way accretion stalls as each successive protoplanet migrates into the region of the disk already picked clean of planetesimals by previous protoplanets.  
Strictly speaking, the mass eliminated at the inner edge is not yet lost since our inner computation boundary is at $r=1$ AU rather than $r=0$.  However, since each of the eliminated protoplanets has already ceased accreting, we are not missing anything interesting by removing them a bit prematurely.
\begin{figure}
\epsscale{0.8}
\plotone{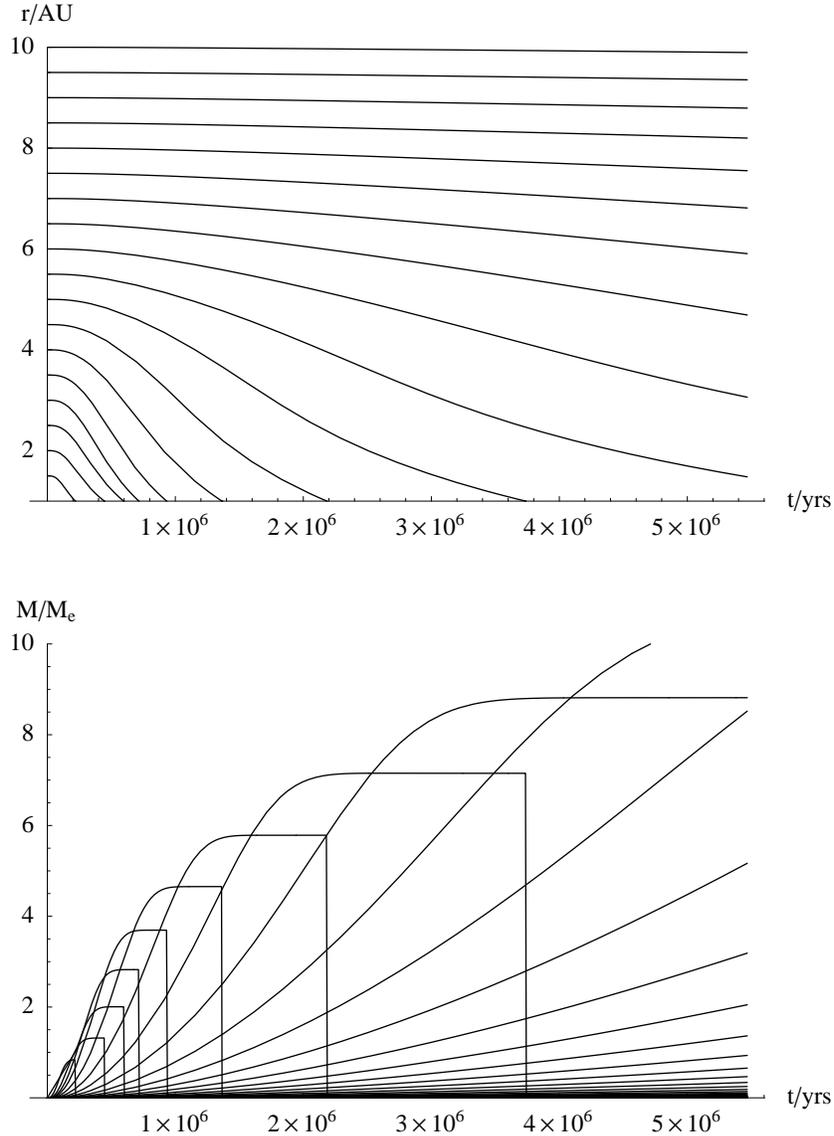}
\caption{The protoplanet evolution paths of Figure \ref{fig: baseline_full_calculation}, projected into the $r$ vs. $t$ plane (TOP) and the $M$ vs. $t$ plane (BOTTOM).  Over the time shown (5.5 Myrs), the innermost eight protoplanets are lost at the inner boundary.  The next three are not lost and grow to 8 M$_\oplus$ or larger.  The remaining protoplanets all remain at masses $< 5$ M$_\oplus$, and undergo relatively little migration.}
\label{fig: timeline_projections}
\epsscale{1}
\end{figure}

\section{Disk properties and core formation}

We now investigate how the model results change with disk properties.  
We already know that the disk must dissipate in order for a window of opportunity for core formation to open; now we wish to quantify the dependence of the model on disk viscosity (as parameterized by $\alpha$).  We also want to examine how the outcome changes with initial disk mass and metallicity.  We can already anticipate the qualitative form of the latter dependency:  Since increasing metallicity increases the planetesimal disk mass, thus tipping the balance away from migration and toward accretion, we expect that for a given set of disk parameters, increasing [Fe/H] will always produce larger protoplanets within a given time.  However, dependence on initial gas disk mass is not a priori obvious.  
Since we take $\Sigma_m \propto \Sigma_g(0) \propto M_d(0)$, the growth rate at $t=0$ goes as $dM/dt \propto \Sigma_g^{2/5} \Sigma_m \propto M_d(0)^{7/5}$.  Meanwhile, the migration rate at $t=0$ goes as $dr/dt \propto \Sigma_g \propto M_d(0)$.  Thus, increasing $M_d(0)$ boosts the {\it initial} advantage of accretion over migration, but as illustrated in Figure \ref{fig: migr_no_disk_evol}, an initial spike in protoplanet mass can be rapidly erased by migration.  Figure \ref{fig: 4line_estimate_4panels} repeats the computation shown in Figure \ref{fig: 4line_estimate_plot} with four different disk masses, keeping $\alpha=10^{-2}$.  We see that initially, $M_{\rm max}$ is set by the isolation mass $M_{\rm iso}$ rather than by $M_{\rm cross}$, and increases with $M_d(0)$.  $M_{\rm cross}$ decreases with disk mass while $M_{\rm iso}$ increases, until $M_{\rm max}$ is set by $M_{\rm cross}$.  For even larger disk masses, $M_{\rm max}$ then remains approximately constant but occurs at ever-larger $r$.  
\begin{figure}
\plotone{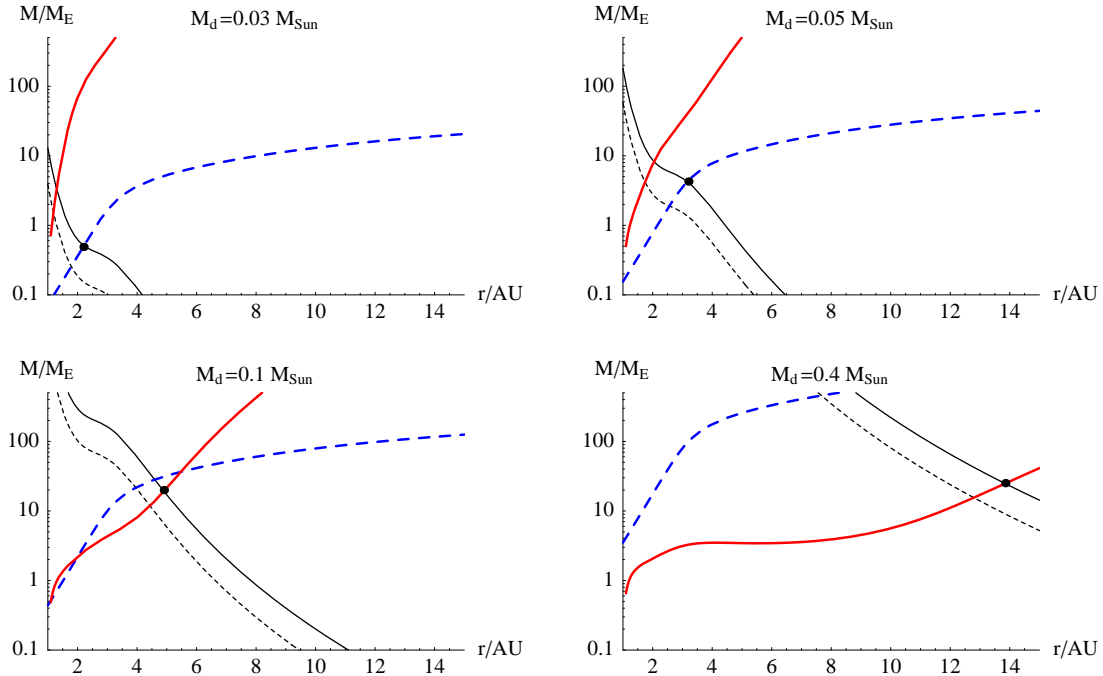}
\caption{The calculation shown in Figure \ref{fig: 4line_estimate_plot}, repeated for disk masses of 0.03, 0.05, 0.1 and 0.4 M$_\odot$.  In each case, a filled circle shows the estimate of $M_{\rm max}$ at $t=t_{\rm 100 AU}$.} 
\label{fig: 4line_estimate_4panels}
\end{figure}

We can rapidly scan a large parameter space using the estimation scheme of \S \ref{sec: timescales}.  For each set of parameters, we can calculate $M_{\rm max}$ (Equation \ref{eq: maximum mass estimate}).  We proceed by varying $M_d(0)$ and $\alpha$, holding all other parameters fixed at the values used for \S \ref{sec: baseline}.  We consider disk masses from 0.01 to 0.4 M$_\odot$, and $\alpha$ from $10^{-4}$ to $10^{-1}$.  We repeat this for [Fe/H]=-0.25, 0, 0.25 and 0.5.  The results are shown in Figure \ref{fig: Md_vs_alpha_vs_metallicity_estimate}.
\begin{figure}
\epsscale{1}
\plotone{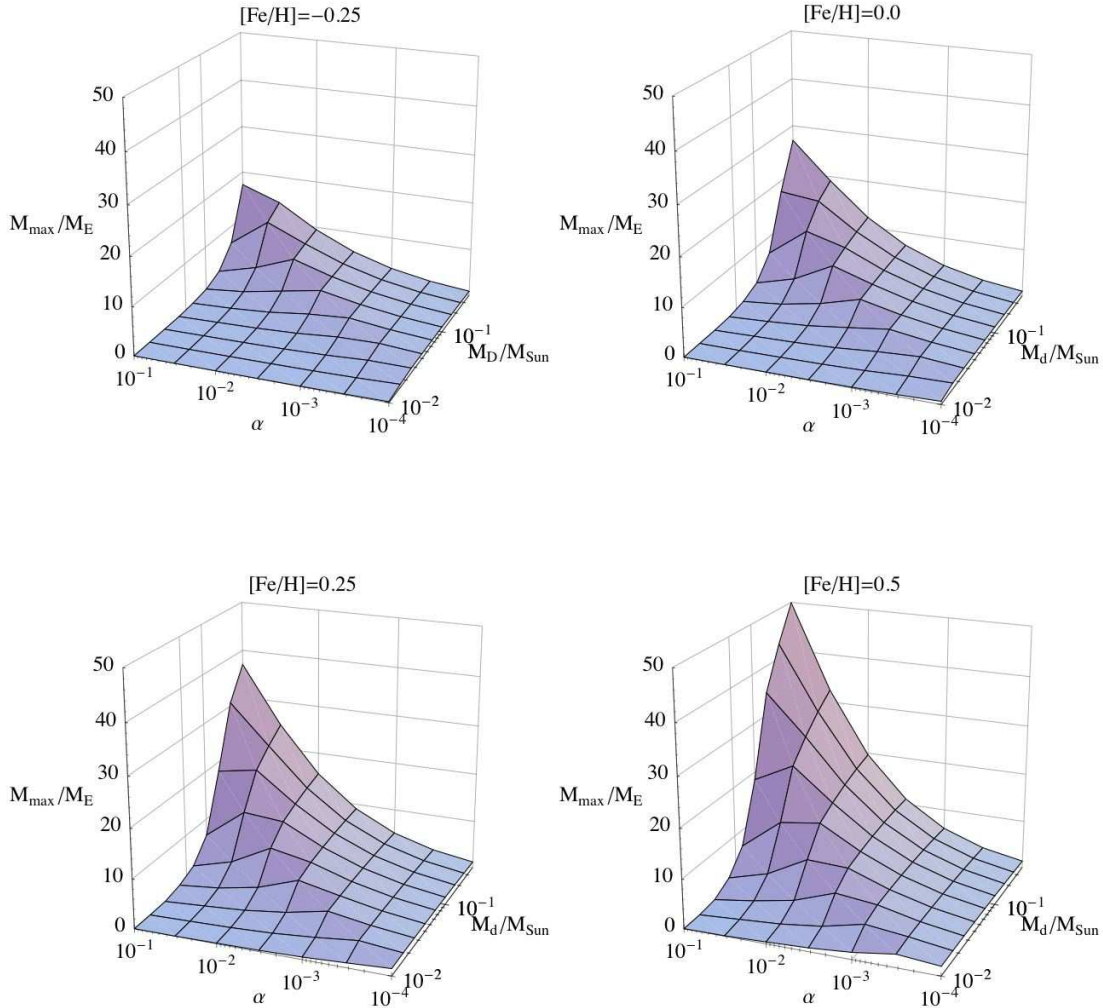}
\caption{The largest protoplanet mass, $M_{\rm max}$ (M$_\oplus$), at the time the gas disk mass is 1 M$_{\rm Jup}$ inside 100 AU, using the estimate of \S \ref{sec: timescales}, for [Fe/H]=-0.25, 0, 0.25 and 0.5.}
\label{fig: Md_vs_alpha_vs_metallicity_estimate}
\end{figure}
As expected, $M_{\rm max}$ increases with  metallicity.  Also, we see that for a given value of the disk mass, there is indeed an``optimal" value of $\alpha$, $\equiv \alpha_{\rm best}$, which produces the largest $M_{\rm max}$.   If $\alpha \ll \alpha_{\rm best}$, the disk does not dissipate fast enough and significant migration takes place, resulting in an outcome similar to those depicted in Figure \ref{fig: migr_no_disk_evol}.  On the other hand, if $\alpha \gg \alpha_{\rm best}$ the gas disk dissipates {\it too} fast, so that by $t=t_{\rm 100\,AU}$, little accretion has occurred.  The value of $\alpha_{\rm best}$ increases with disk mass and metallicity.  Finally, we see the dependence on disk mass indicated by Figure \ref{fig: 4line_estimate_4panels}: for a given value of $\alpha$, $M_{\rm max}$ initially increases with $M_d(0)$, then levels off.  The lower $\alpha$, the lower the disk mass at which $M_{\rm max}$ stops increasing.

We now calculate a grid of full models as in \S \ref{sec: baseline} for the same parameter range as above, in each case extracting $M_{\rm max}$ at $t=t_{\rm 100\,AU}$;  the results are shown in Fig \ref{fig: Md_vs_alpha_vs_metallicity}.  Relative to Figure \ref{fig: Md_vs_alpha_vs_metallicity_estimate}, $M_{\rm max}$ is less strongly peaked toward large disk mass and high viscosity.   At low viscosity, rather than leveling off, $M_{\rm max}$ increases monotonically with initial disk mass for a given $\alpha$.  
Our simple estimate nevertheless fares relatively well in qualitatively reproducing the $M_{\rm max}-M_d(0)-\alpha-{\rm [Fe/H]}$ relationship, and over the plotted range gets the maxima of the $M_{\rm max}$ surfaces right within a factor of $\la 1.5$.  

\begin{figure}
\plotone{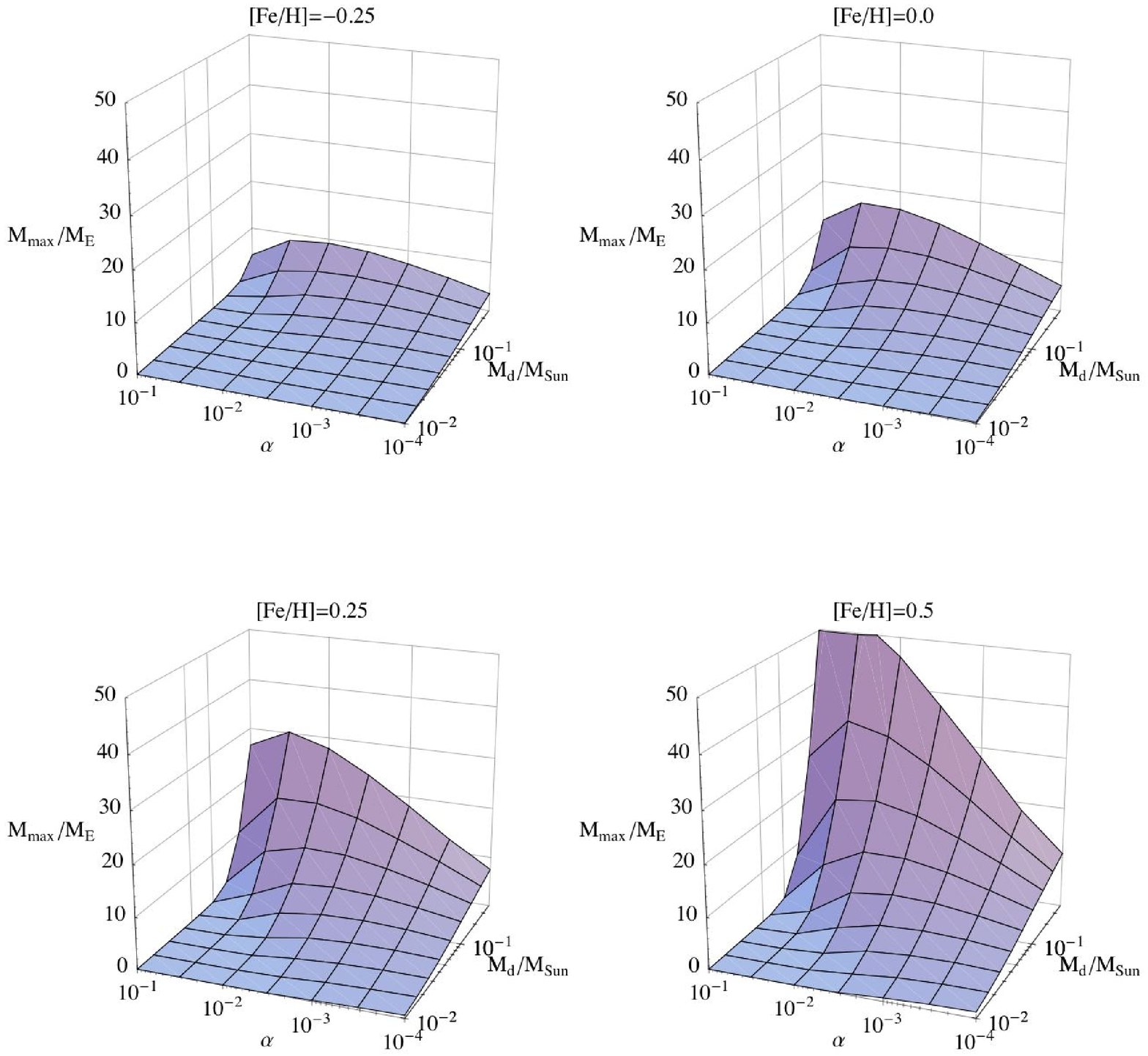}
\caption{The largest protoplanet mass, $M_{\rm max}$, at the time the gas disk mass is 1 M$_{\rm Jup}$ inside 100 AU, computed using the full model, for [Fe/H]=-0.25,0, 0.25, 0.5.  }
\label{fig: Md_vs_alpha_vs_metallicity}
\end{figure}

From Figure \ref{fig: Md_vs_alpha_vs_metallicity}, we see that a sub-Solar metallicity disk, [Fe/H]=-0.25, falls just short of producing $M_{\rm max}=10$ M$_\oplus$, even with an initial disk mass of $M_d(0)=0.4$ M$_\odot$.  A Solar-metallicity disk produces $M_{\rm max}=10$ M$_\oplus$ if it starts out with $M_d(0) \ga 0.3$ M$_\odot$, provided also $\alpha \sim 10^{-2}$.
With a super-Solar metallicity of [Fe/H]=0.25, a disk mass slightly above $0.1$ M$_\odot$ can produce $M_{\rm max}=10$ M$_\oplus$ if, again, $\alpha \sim 10^{-2}$.  Finally, with a very high metallicity of [Fe/H]=0.5, a disk mass $\la 0.1$ M$_\odot$ suffices to make $M_{\rm max}=10$ M$_\oplus$, if also $10^{-3} \la \alpha \la 10^{-2}$.  

\section{Discussion}
\label{sec: discussion}

We have demonstrated that bodies $\sim 10$ M$_\oplus$ in mass, large enough to initiate the formation of gas giants by nucleated instability, can grow in a viscously evolving gas disk even if they are subject to Type I migration.  Though the accretion timescale is significantly longer than the migration timescale in a gas disk that is young and massive, $M_{\rm d} \sim 10^{-1}$ M$_\odot$, the times become comparable when $M_d \sim 10^{-2} - 10^{-3}$ M$_\odot$;
in other words, when there are only a few Jupiter masses of gas left in the disk.  At this point, a protoplanet has a chance of growing to core mass before spiraling into the star.  In this picture, then, the formation of giant planet cores does not take place until a window of opportunity opens in the latter part of the gas disk's lifetime.  

At the same time, though, our findings suggest that an era of successful core formation is by no means inevitable in a protoplanetary disk.  The accretion of 10 M$_\oplus$ or larger bodies necessitates an initial disk mass $M_d(0) > 0.1$ M$_\odot$ at Solar metallicity, and not much less than $0.1$ M$_\odot$ at [Fe/H]=0.5.  A lower-metallicity disk having [Fe/H]=-0.25 does not produce 10 M$_\oplus$ bodies unless it is very massive, $M_d(0) > 0.4$ M$_\odot$.
Furthermore, at a given disk mass the largest protoplanets occur only at a particular viscosity.  Too much lower, and the disk ``overstays its welcome", removing too much mass in abortive cores from the disk.  Too much higher, and the disk dissipates too rapidly, without core accretion having had a chance to get very far.  However, assuming disks to be adequately described by an alpha parameterization of viscosity, the values we obtain for $\alpha_{\rm best}$, $\sim 10^{-2}$ and as low as $10^{-3}$, agree well with values of $\alpha$ obtained from fits to observed disk lifetimes and accretion rates 
\citep{1998ApJ...495..385H}.
We have not considered disk photoevaporation 
\citep{1993Icar..106...92S,2003ApJ...585L.143M}.
In a disk where this exists as an additional sink of disk gas, $\alpha_{\rm best}$ will be correspondingly lower than it would be for pure accretion onto the star.  We defer a detailed examination to future work.  

It has been suggested 
\citep[e.g.][]{2002A&A...394..241T}
that protostellar disks produce numerous gas giants, but that inward migration in an annular gap (Type II migration) removes many of them as the disk accretes onto the star, leaving only the last few survivors to populate the mature system.  The model described here is analogous to this picture, except that the culling of giant planets occurs already at the core stage.  
In fact, our analysis suggests that full-sized gap-opening gas giants do not have a chance to grow until the gas disk is quite tenuous.  Since the runaway accretion of the gas envelope can proceed at a rate $\ga 10^{-5}$ M$_{\rm Jup}$/yr
\citep{1996Icar..124...62P}, 
while by this time the disk accretion rate is $\sim 10^{-6} - 10^{-5}$ M$_{\rm Jup}$/yr (\S \ref{sec: timescales}), a significant fraction of the disk's accretion likely goes into feeding the planet rather than pushing it inward.  Once the planet grows massive enough ($\sim$1 M$_{\rm Jup}$) to open a deep gap and enter the Type II regime, its mass is comparable to that of the remaining disk, 
so that significant Type II migration is unlikely to occur.  Therefore in our scenario it becomes Type I rather than Type II migration which plays the primary role in determining the radial distribution of gas giants.  

The above argument notwithstanding, a detailed examination of the endgame of giant planet formation in a tenuous gas disk is an important area for future work.  Is our adopted criterion of a Jupiter mass inside 100 AU a sufficient condition for gas giant formation?  \cite{1996Icar..124...62P}
find that even after a core grows to $\sim 10$ M$_\oplus$, a plateau of slow gas accretion can exist for many millions of years before runaway gas accretion supplies the massive jovian atmosphere.  If this is the case, lots of extra gas will be required in the disk.  However, these model used the interstellar value for grain opacity; reducing this to account for settling and coagulation of dust in a protostellar disk can substantially reduce the length of the plateau, as shown by  
\cite{2005Icar..179..415H}.
 The authors also demonstrate that cutting off planetesimal accretion onto a core speeds the onset of runaway gas accretion, by reducing the rate of energy deposition into the envelope.  In fact, migration does exactly that, as illustrated in Figure \ref{fig: timeline_projections}: As protoplanets move into parts of the disk already picked clean of planetesimals by their inner neighbors, they cease accreting.  
  
Once the first gap-opening planet does form, it will act as a barrier in the disk to faster migrators originating from outside its orbit 
\citep{2005ApJ...626.1033T},
thus the formation of subsequent planets---including gas giants if there is enough gas left in the disk---becomes easier.  Before that, however, a significant total mass in failed cores may migrate to the central star.   Recent discoveries of short-period exoplanets with a minimum mass comparable to that of Neptune orbiting GJ 436 
\citep{2004ApJ...617..580B}
and 55 Cancri 
\citep{2004ApJ...614L..81M}
may represent such would-be cores, which survived either through luck (the gas disk was removed just in time) or through some migration-stopping mechanism near the star.  

The observed correlation between host star metallicity and planet occurrence rate 
\citep{1997MNRAS.285..403G,2005ApJ...622.1102F}
is seen as evidence that gas giants begin their existence as solid cores, since more solid material allows a disk to form larger solid bodies, and to do so faster.  The model of concurrent core migration and accretion developed here provides an even stronger theoretical basis for a planet-metallicity correlation:  The higher the ratio of solids in the disk, the better accretion does relative to migration, all other things being equal.  

Finally, as alluded to in \S \ref{sec: intro}, we have considered here what is essentially the worst-case scenario of Type I migration.  
Details of the disk structure, not considered in our simple model, may act to significantly reduce migration.  The MHD turbulence providing the disk's viscosity may simultaneously produce enough fluctations in the gas density (and thus in the disk torques) to give Type I migration the character of a random walk in radius 
\citep{2004ApJ...608..489L,2004MNRAS.350..849N},
 thus enabling at least a tail of the protoplanet population to survive even in a massive disk.  Persistent rather than random jumps in disk properties may do even better.  
\cite{2004ApJ...606..520M}
 show that at the locations of opacity transitions, Type I migration may slow down by an order of magnitude.  The density jump associated with the outer edge of a dead zone 
\citep{1996ApJ...457..355G}
may stop Type I migration altogether (Matsumura, Thommes \& Pudritz, in preparation).   Help may also come on the other side of the migration-accretion contest.  As mentioned in \S \ref{sec: rates}, our simple accretion rate estimate neglects (i) planetesimal migration by gas drag, (ii) planetesimal fragmentation, (iii) the growth of gas envelopes on protoplanets, (iv) enhanced accretion before planetesimal random velocities catch up to their equilbrium values, and (v) the contribution to protoplanet growth by mergers among protoplanets.  Effects (ii)-(v) enhance accretion, and although (i) by itself reduces accretion efficiency \citep{2003Icar..161..431T}, this is largely offset \citep{2003Icar..166...46I,2006Icar..180..496C} by (ii) and (iii): smaller planetesimals experience stronger gas drag, which increases their migration,  but also increases the efficiency with which they are captured by protoplanet atmospheres.  
Thus, our estimate of the growth rate is likely conservative.  Also, we have assumed that the ratio between the solid and gaseous components throughout the protostellar disk at $t=0$ is just given by the metallicity; in reality, large local density enhancements in the solids may develop 
\citep{1988Icar...75..146S,2004ApJ...614..490C},
giving a correspondingly large boost to core accretion at those locations.  

In short, there is an abundance of candidates for countering the effect of Type I migration.  Our purpose here has been to demonstrate that even in the absence of any such mechanisms, Type I migration is far from being an insurmountable obstacle to the formation of gas giants by core accretion.     

\section{Summary}

The formation of giant planets by core accretion requires the growth of solids bodies $\sim 10$ M$_\oplus$ in mass.  Bodies of this mass are well below the threshold for gap formation, and will thus migrate inward relative to the gas due to the imbalance between inner and outer planet-disk torques (Type I migration).  In a gas disk with mass of a few $\times 10^{-2}$ M$_\odot$ or greater, the Type I migration timescale of core-mass bodies is shorter than their formation time, so that growing protoplanets plunge into the star before they have a chance to become cores.  
We have demonstrated here that cores can form nevertheless, and that it is the ``late bloomers" which are actually favored.  Protoplanets which grow large early suffer strong migration and are lost.  Those which grow later, however, do so in a more tenuous gas disk, with a correspondingly longer migration time.  A window of opportunity can thus open between the time when growth and migration timescales become comparable, and the time when the disk no longer contains enough gas to furnish a jovian atmosphere.  The size of this window increases with metallicity and initial disk mass.  

\acknowledgements{We thank Doug Lin for numerous stimulating and informative discussions on this topic, Sarah Nickerson for assistance with the MATLAB version of the computations, and the referee for valuable comments and suggestions.  This work was supported by NSERC of Canada.}

\end{document}